\documentstyle[12pt]{article}
\begin{document}
\begin{center}
{\large\bf A simple entanglement measure for
multipartite pure states}
\vskip .6cm {\normalsize Feng Pan$^{1,2}$,
Dan Liu$^{1}$, Guoying Lu$^{1}$, and J. P. Draayer$^{2}$ }
\vskip .1cm {\small
$^{1}$Department of Physics, Liaoning Normal University, Dalian
116029, P. R. China\vskip .2cm
$^{2}$Department of Physics and Astronomy,
Louisiana State University, Baton Rouge, LA 70803-4001}
\vskip 1cm
\end{center}

\abstract
A simple entanglement measure for multipartite pure states is
formulated based on the partial entropy of a series of reduced
density matrices. Use of the proposed new measure to distinguish
disentangled, partially entangled, and maximally entangled
multipartite pure states is illustrated.



\newpage

Entanglement plays an important role in the theory of quantum
information and quantum computation.$^{[1,2]}$  A major challenge
that remains is how to define good measures of entanglement since
simple measures that classify and quantify entanglement of a given
state should enhance our understanding of the phenomenon. Although,
many measures of entanglement have been proposed,$^{[3-26]}$ most
involve extremizations that are difficult to manage
analytically.$^{[8]}$

There has been a lot of work on multipartite entanglement.
For example, Bennett {\it et al} in [17] introduced exact
and asymptotic measures for multipartite pure state
entanglement, in which a minimal reversible entanglement generating
set (MREGS) was defined. In [18], reversibility of local transformations
of multipartite entanglement was studied.
Relations between tripartite pure state entanglement and additivity
properties of the bipartite relative entropy of entanglement
were established in [19]. Upper and lower bounds to the relative
entropy of entanglement of multi-party systems in terms of the
bi-partite entanglements of formation and
distillation and entropies of various subsystems were
discussed in [20]. Recently, the structure of a reversible
entanglement generating set for three-particle states
were investigated in [21].
In the connection with the logarithmic negativity discussed
in [22], an operational interpretation of the logarithmic
negativity has been found.$^{[23]}$  All these works help
us to get better understandings of the multipartite
entanglement.

A good definition of an entanglement measure can be used to
distinguish entangled, partially entangled, and disentangled
states, and this in turn should be useful in understanding the
extent those particles are entangled and how many ways a
multipartite system can be entangled. For a bipartite pure state,
the problem has been solved.$^{[13]}$ In this case, an entanglement
measure can be defined in terms of the von Neumann entropy. However,
the problem still remains open for a system with more than three
particles. The situation becomes more difficult and unclear for
mixed states. In the following, we will concentrated on multipartite
pure states, for which, as for the spin-${1\over{2}}$ case, there
are two degrees of freedom with $\sigma=0$ or $1$ for each particle.

For a system of $N$ such identical particles, any
wavefunction $\vert\Psi\rangle$ can be expanded in terms of
basis vectors
$\vert \sigma_{1},\sigma_{2},\cdots ,\sigma_{N} \rangle$
in the tensor product space
$\left(V_{2}\bigotimes\right)^{N}$
as

$$\vert\Psi\rangle=\sum_{\sigma_{1}\cdots \sigma_{N}}
C_{\sigma_{1}\cdots \sigma_{N}}
\vert \sigma_{1},\cdots,\sigma_{N}\rangle,\eqno(1)$$
where $\sigma_{i}= 0$ or $1$ for $1\leq i\leq N$ and
$C_{\sigma_{1}\cdots \sigma_{N}}$
is the normalized expansion coefficient. The corresponding
density matrix is

$${\rho}_{\Psi}=\vert\Psi\rangle\langle\Psi\vert.\eqno(2)$$

\noindent Let $a_{i\sigma}^{\dagger}$ ($a_{i\sigma}$)
with $i=1,2,\cdots,N$, be particle
creation (annihilation) operators that satisfy

$$[a_{i\sigma},a_{j\sigma^{\prime}}^{\dagger}]_{\pm}\equiv
a_{i\sigma}a_{j\sigma^{\prime}}^{\dagger}\pm
a_{j\sigma^{\prime}}^{\dagger}a_{i\sigma}=
\delta_{ij}\delta_{\sigma\sigma^{\prime}},\eqno(3a)$$

$$[a_{i\sigma},a_{j\sigma^{\prime}}]_{\pm}=
[a^{\dagger}_{i\sigma},a_{j\sigma^{\prime}}^{\dagger}]_{\pm}=0,
\eqno(3b)$$
for spin-$1\over{2}$ fermions or 2-component bosons.
The wavefunction $\vert\Psi\rangle$ can be expressed as

$$\vert\Psi\rangle=\sum_{\sigma_{1}\cdots\sigma_{N}}
C_{\sigma_{1}\cdots \sigma_{N}}a_{1\sigma_{1}}^{\dagger}
a_{2\sigma_{2}}^{\dagger}\cdots a_{N\sigma_{N}}^{\dagger}
\vert 0\rangle,\eqno(4)$$
where $\vert 0\rangle$ is the vacuum state.
Under the replacement $a_{i\sigma_{i}}^{\dagger}\rightarrow
X_{i\sigma_{i}}$, where $X_{i\sigma_{i}}$ is simply a symbol,
the operator form in front of the vacuum state on the
right-hand-side of Eq. (4) becomes a homogeneous polynomial
of degree $N$ in terms of the $\{{\bf X}_{i}\}$,

$$F_{C}({\bf X}_{1},\cdots,{\bf X}_{N})
=\sum_{\sigma_{1}\cdots\sigma_{N}}
C_{\sigma_{1}\cdots \sigma_{N}}X_{1\sigma_{1}}
\cdots X_{N\sigma_{N}}.\eqno(5)$$
It should be understood that ${\bf X}_{i}$ is
a two-value symbol with ${\bf X}_{i}=X_{i1}$ and $X_{i0}$.
An alternative definition of entangled states can be
stated as follows: The state $\vert\Psi\rangle$ is
an $N$-particle entangled state if the corresponding
polynomials $F_{C}({\bf X}_{1},\cdots,{\bf X}_{N})$
on complex field ${\bf\cal C}$
cannot be factorized into the following form

$$F_{C}({\bf X}_{1},\cdots,{\bf X}_{N})
=F_{A}({\bf X}_{i_{1}},\cdots,{\bf X}_{i_{m}})
F_{B}({\bf X}_{i_{m+1}},\cdots,{\bf X}_{i_{N}})\eqno(6)$$
for $1\leq m\leq N-1$, where
$\{i_{1}\neq i_{2}\neq\cdots\neq i_{N}\}$ can be in any ordering
of $\{1,2,\cdots,N\}$. Otherwise the state $\vert\Psi\rangle$
is not an $N$-particle entangled state.
  The state $\vert\Psi\rangle$ given in (4) is disentangled
(separable) if the polynomials $F_{C}$ can be factorized into
a product of monomial of ${\bf X}_{i}$ as
$\prod_{i=1}^{N}F_{A_{i}}({\bf X}_{i})$. In other cases,
the state is partially entangled.

For $N=2$, a criterion for distinguishing whether
a homogeneous polynomial is factorizable can be
established by using the von Neumann entropy of the
reduced density matrix. Furthermore, the degree of
entanglement can be quantified by the von Neumann
entropy with

$$S_{\Psi}=-{\rm Tr}(\left(\rho_{\Psi}\right)_{i}{\rm Log}_{2}
\left(\rho_{\Psi}\right)_{i})
=-{1\over{2}}\left({\rm Tr}(\left(\rho_{\Psi}\right)_{1}{\rm Log}_{2}
\left(\rho_{\Psi}\right)_{1})+
{\rm Tr}(\left(\rho_{\Psi}\right)_{2}{\rm Log}_{2}
\left(\rho_{\Psi}\right)_{2})\right)
,\eqno(7)$$
where $i=1$ or $2$, and $\left(\rho_{\Psi}\right)_{i}$ ($i=1$ or
$2$) is the reduced density
matrix with particle $2$ or $1$, respectively, traced out.
This definition and the correspondence between the
factorizable (non-factorizable) case of (5) and a disentangled
(entangled) state given in (4) is well-known,
which provides with a clear  quantification of entanglement
for bipartite pure states.
A state is separable if $S_{\Psi}=0$, entangled if $S_{\Psi}\neq 0$,
and maximally entangled if $S_{\Psi}=1$. In (7), we have
used the fact that
$\left(\rho_{\Psi}\right)_{1}=\left(\rho_{\Psi}\right)_{2}$.

However, there will be many new features for $N\geq 3$.
Let $\left(\rho_{\Psi}\right)_{(12\cdots N-1)}$ be the
reduced density matrix with the $N$-th particle traced out.
There is a series of reduced density matrices with $N-1$
particles,

$$\{Q^{N-1}_{\omega}\left(\rho_{\Psi}\right)
_{(12\cdots N-1)}\},\eqno(8)$$
where $Q^{N-1}_{\omega}$ is the left coset representative
of the factor group $S_{N}/(S_{N-1}\otimes S_{1})$, in which
$S_{k}$ is the permutation group, and
$\omega$ is the normal ordered sequences.$^{[24]}$
Let $g_{i}$ ($i=1,2,\cdots,N-1$) be generators of $S_{N}$, which
are  adjacent permutation of the $i$-th and $(i+1)$-th
particles. When $N=3$, one has $\{ Q^{2}_{1}=e,
Q^{2}_{2}=g_{2}, Q^{2}_{3}=g_{1}g_{2}\}$. Thus, one gets
three two-particle reduced density matrices
$\left(\rho_{\Psi}\right)_{(12)}$,
$\left(\rho_{\Psi}\right)_{(13)}$, and
$\left(\rho_{\Psi}\right)_{(23)}$ according to (8).
Consequently, there will be a series of reduced density
matrices with $N-2$, $N-3$,$\cdots$, $1$ particle(s),
$\{Q^{N-2}_{\omega_{N-2}}\left(\rho_{\Psi}\right)_{(12\cdots
N-2)}\}$, $\{Q^{N-3}_{\omega_{N-3}}\left(\rho_{\Psi}\right)_{(12\cdots
N-3)}\}$,$\cdots$,
$\{Q^{1}_{\omega_{1}}\left(\rho_{\Psi}\right)_{(1)}\}$,
where $Q^{N-k}_{\omega_{N-k}}$ is the left coset representative
of the factor group $S_{N}/(S_{N-k}\otimes S_{{k}})$.
For $N=3$, a complete set of reduced density matrices
is $\{ \left(\rho_{\Psi}\right)_{(12)},
\left(\rho_{\Psi}\right)_{(13)},
\left(\rho_{\Psi}\right)_{(23)}$,
${\left(\rho_{\Psi}\right)_{(1)},
\left(\rho_{\Psi}\right)_{(2)},
\left(\rho_{\Psi}\right)_{(3)}\}}.$
It can be shown that the state $\vert\Psi\rangle$ is
not a genuine  $N$-particle entangled state if the
von Neumann entropy defined in terms
of one of the reduced density matrices in the series
$\{Q^{N-k}_{\omega_{N-k}}\left(\rho_{\Psi}\right)_{(12\cdots
N-k)}\}$ ($k=1,2,\cdots,N-1$) is zero because
the corresponding homogeneous polynomial (5) is, at least,
partially factorizable.
Furthermore, unlike the $N=2$ case,
it can be verified that values of the von Neumann entropy
of reduced matrices for $N-k$ particles with $k=1,2,\cdots,
N-1$, are not the same for fixed $k$. For example, generally,
$\left(\rho_{\Psi}\right)_{(12)}\neq
\left(\rho_{\Psi}\right)_{(23)}\neq
\left(\rho_{\Psi}\right)_{(13)}$, and
$\left(\rho_{\Psi}\right)_{(1)}\neq
\left(\rho_{\Psi}\right)_{(2)}\neq
\left(\rho_{\Psi}\right)_{(3)}$.
In addition, it will be shown later
that the maximal entropy calculated
from $\{Q^{1}_{\omega}\left(\rho_{\Psi}\right)_{(1)}\}$
may be less than $1$ when $N\geq 3$.

Based on the above observations, we can defined
a measure of genuine $N$-particle entanglement  $\eta^{(N)}_{\Psi}$
as follows:

$$\eta^{(N)}_{\Psi}=
\left\{\matrix{{1\over{N}}\sum^{N}_{i=1}
{S_{(i)}}
~~{\rm if}~~Q^{N-k}_{\omega_{N-k}}S_{(12\cdots N-k)}\neq 0
~~\forall~~\omega_{N-k}~~{\rm with}~~1\leq k\leq N-1,
\cr
{}\cr
0~~~~~{\rm otherwise,}~~~
~~~~~~~~~~~~~~~~~~~~~~~~~
~~~~~~~~~~~~~~~~~~~~~
~~~~~~~~~~~~~~~~~~~~\cr
}\right.\eqno(9)$$
where

$$S_{(12\cdots N-k)}=-{\rm Tr}\left(
(\rho_{\Psi})_{(12\cdots N-k)}{\rm Log}_{2}
(\rho_{\Psi})_{(12\cdots N-k)}\right)\eqno(10)$$
is the partial von Neumann entropy with the $k$ particles
traced out.
The state $\vert\Psi\rangle$ is, at least, partially
separable when one of the values of partial entropy
$\{Q^{N-k}_{\omega_{N-k}}S_{(12\cdots N-k)}\}$ is zero.
In such case, the corresponding state is not a
genuine $N$-particle entangled state.
Otherwise, we can quantify the measure using the average
one-particle reduced entropy defined in (9).

It is clear that (9) is zero for separable states.
Furthermore, the entanglement measure
should be invariant under local unitary transformations,
and its expectation should not increase under
local operations and classical communication
(LOCC).  In order to prove (9) satisfying
the above requirements, we use the conclusions
made in [17]. As has been noted in [17], partial
entropies have the nice property that for pure
states their average does not increase under
LOCC. The entanglement measure (9) is
defined in terms of the average one-particle
reduced entropy. Therefore, its value should
also not increase under LOCC. In addition,
since the measure (9) is defined in terms
of the average one-particle reduced entropy, its
value should also be invariant under any
local unitary transformation.

However, Eq. (9) does not tell us how many particles
are disentangled from one another and
how many of them are still entangled,
when some of the values of the partial entropy $\{
Q^{N-k}_{\omega_{N-k}}S_{(12\cdots N-k)}\}$ are zero.
Actually, one need
to calculate all values of the partial entropy
$\{Q^{N-k}_{\omega_{N-k}}S_{(12\cdots N-k)}\}$
to get the full picture.
In the following, as an example, we will apply the above
definition of the entanglement measure given by (9) for the
$3$-particle case. For $N=3$, the complete
basis space is eight dimensional ($2^N = 8$),
of which the basis vectors are denoted as
$\{\vert 1\rangle=\vert 000\rangle, \vert 2\rangle=\vert 110\rangle,
\vert 3\rangle=\vert 101\rangle,\vert 4\rangle=\vert 011\rangle,
\vert 5\rangle=\vert 111\rangle,\vert 6\rangle=\vert 001\rangle,
\vert 7\rangle=\vert 010\rangle,\vert 8\rangle=\vert
100\rangle\}$. A general $3$-particle state $\vert\Psi\rangle$
can be expanded in terms of these basis vectors with at most
$8$ terms. As a simple example, we assume a $3$-particle state
$\vert\Psi\rangle$ has three non-zero terms. There are $56$
possible three-term linear combinations ($(^8_3) =56$) of
these $8$ basis vectors as listed in Table 1, with $24$
combinations being states with $2$-particle entangled
and disentangled with another one. Therefore, those $24$
states are partially entangled, and not genuine $3$-particle
entangled states. The remaining $32$ such combinations are
genuine $3$-particle entangled states. In order to verify the
effectiveness of the definition (9), we calculate a series of
reduced density matrices for two cases with
$\vert\Psi\rangle=\vert127\rangle$ and
$\vert123\rangle$ given in the Table 1.
In the case of $\vert127\rangle$, the wavefunction can be
written as

$$\vert\Psi\rangle=\alpha\vert 000\rangle+\beta\vert110\rangle
+\gamma\vert010\rangle,\eqno(11)$$
where $\alpha$, $\beta$, and $\gamma$ are nonzero complex numbers
satisfying the normalization condition.
The corresponding diagonalized reduced density matrices are

$$(\rho)_{(12)}=\left(\matrix{0\cr{}  &1\cr}\right),~~
(\rho)_{(13)}=(\rho)_{(23)}=
\left(\matrix{{1\over{2}}(1-\sqrt{1-4
\vert\alpha\vert^{2}\vert\beta\vert^{2}})\cr{}&
{1\over{2}}(1+\sqrt{1-4
\vert\alpha\vert^{2}\vert\beta\vert^{2}})\cr}\right),$$

$$(\rho)_{(1)}=(\rho)_{(2)}=
(\rho)_{(13)},~~(\rho)_{(3)}=(\rho)_{(12)}.\eqno(12)$$
Therefore, the corresponding values of partial entropy
are $S_{(12)}=0$, $S_{(13)}=S_{(23)}\neq 0$,
$S_{(1)}=S_{(2)}\neq 0$, and $S_{(3)}=0$.
The values of entropy $S_{(3)}=S_{(12)}=0$ indicate that in
this case particle $3$ is disentangled from particle
$1$ and $2$, while the values of entropy $S_{(1)}=S_{(2)}\neq
0$ indicate that particle $1$ and $2$ are still entangled.
According to definition (9), therefore, the $3$-particle state
$\vert127\rangle$ is not a genuine $3$-particle entangled state.
When $\vert\Psi\rangle=\vert123\rangle$, the corresponding
diagonalized reduced density matrices are
$$(\rho)_{(3)}=(\rho)_{(12)}=\left(\matrix{\vert\alpha\vert^{2}+
\vert\beta\vert^{2}\cr{}  &\vert\gamma\vert^{2}\cr}\right),~~
(\rho)_{(2)}=(\rho)_{(13)}=\left(\matrix{\vert\beta\vert^{2}
\cr{}  &\vert\vert\alpha\vert^{2}+\gamma\vert^{2}\cr}\right),$$
$$(\rho)_{(1)}=(\rho)_{(23)}=
\left(\matrix{\vert\alpha\vert^{2}\cr{}&
\vert\beta\vert^{2}+\vert\gamma\vert^{2}\cr}\right).\eqno(13)$$
Hence, the corresponding values of partial entropy are all
non-zero, which indicate that the state $\vert123\rangle$ is
a genuine $3$-particle entangled state. In this case, the values
of reduced entropy $(S)_{(i)}$ for $i=1,2,3$ are not the same
in general.
To maximize (9) with the results given in (13) and the constraint
$\vert\alpha\vert^{2}+\vert\beta\vert^{2}+
\vert\gamma\vert^{2}=1$, one finds that $S_{(i)}=0.918296$ for
$i=1,2,3$. Up to a phase factor, the corresponding coefficients
are
$\vert\alpha\vert=\vert\beta\vert=\vert\gamma\vert={1\over{\sqrt{3}}}$,
which gives the maximally entangled $3$-particle state with $3$
terms. Actually, this state belongs to the
$W$-state family.$^{[25]}$ The reduced entropy $S_{(i)}$ is different
from that of two-term GHZ-state case,$^{[26]}$ in which $S_{(i)}=1$.
It has been verified that the definition (9)
is indeed invariant under any local unitary transformation.
This procedure enabled us to analyze all possible $3$-particle
entangled pure states with at most $8$ terms. The detailed results will
be reported elsewhere. The generalization to multipartite
entangled pure states is straightforward.

In summary, we have formulated a simple
entanglement measure for multipartite pure states
based on partial entropy of a series of
reduced density matrices. The new definition
seems suitable to distinguish from disentangled,
partially entangled, and maximally entangled multipartite
pure states. However, entanglement measure of a multipartite mixed
state is much more difficult to be defined than that of
a multipartite pure state studied in this paper.
Much work remain to be done for  multipartite mixed
states.

\section*{Acknowledgments}
This work was supported by the U. S. National Science Foundation, Grants
No. 0140300; the Southeastern Universitites Research Association; the
Natural Science Foundation of China, Grant No. 10175031; and the Natural
Science Foundation of Liaoning Province, Grant No. 2001101053.

\def\HT{\bf\relax}
\def\REF#1{\small\par\hangindent\parindent\indent\llap{#1\enspace}\ignorespaces}

\section*{Reference}

\noindent \REF{[1]} M. A. Nielsen and I. L. Chuang,
Quantum Computation and Quantum Information
(Cambridge University Press, Cambridge, England, 2000)

\REF{[2]} C. H. Bennett and D. P. DiVincenzo, Nature (London) {\bf 404},
247 (2000)

\REF{[3]} S. M. Barnett and S. D. J. Phoenix, Phys. Rev. {\bf A48},
R5 (1993)

\REF{[4]} C. H. Bennett, D. P. DiVincenzo, J. A. Smolin, and W. K. Wootters,
  Phys. Rev. {\bf A54}, 3824 (1996)

\REF{[5]} V. Vedral, M. B. Plenio, M. A. Rippin, and P. L. Knight,
Phys. Rev. Lett. {\bf 78}, 2275 (1997)

\REF{[6]} S. Hill and W. K. Woottes, Phys. Rev. Lett. {\bf 78}, 5022 (1998)

\REF{[7]} L. K. Grover, Phys. Rev. Lett. {\bf 79}, 325 (1997)

\REF{[8]} W. K. Wootters, Phys. Rev. Lett. {\bf 80}, 2245 (1998)

\REF{[9]} V. Vedral and M. B. Plenio, Phys. Rev. {\bf A57}, 1619 (1998)

\REF{[10]} E. M. Rains, Phys. Rev. {\bf A60}, 179 (1999)

\REF{[11]} P. Horodecki, M. Horodecki, and R. Horodecki,
Phys. Rev. Lett. {\bf 85}, 433 (2000)

\REF{[12]} V. Coffman, J. Kundu, and W. K. Wootters,
Phys. Rev. {\bf A61}, 052306 (2000)

\REF{[13]} S. Virmani and M. B. Pleino, Phys. Lett. {\bf A268},
  31 (2000)

\REF{[14]} A. Wong and N. Christensen, Phys. Rev. {\bf A63},
044301 (2001)

\REF{[15]} S. Popescu, D. Rohrlich, J. A. Smolin, and
A. V. Thapliyal, Phys. Rev. {\bf A63},
012308 (2001)

\REF{[16]} A. F. Abouraddy, B. E. A. Saleh, A. V. Sergienko, and
M. C. Teich, Phys. Rev. {\bf A64},
050101 (2001)

\REF{[17]} C. H. Bennett, S. Popescu, D. Rohrlich, J. A. Smolin, and A. V.
Thapliya,  Phys. Rev. {\bf A63}, 012307 (2001)

\REF{[18]} N. Linden, S. Popescu, B. Schumacher, and M. Westmoreland, quant-ph/9912039

\REF{[19]} E.F. Galvao, M.B. Plenio, and S. Virmani,
J. Phys. {\bf A33}, 8809 (2000)

\REF{[20]} M.B. Plenio and V. Vedral, J. Phys. {\bf A34}, 6997 (2001)

\REF{[21]} A. Acin, G. Vidal, and  J. I. Cirac, Quant. Infor. Comput. {\bf 3}, 55 (2003)

\REF{[22]} G. Vidal and R. F. Werner, Phys. Rev. {\bf A65},
032314 (2002)

\REF{[23]} K. Audenaert, M. B. Plenio, and J. Eisert, Phys. Rev.
Lett.  {\bf 90}, 027901 (2003)

\REF{[24]} J. Q. Chen, Group Representation Theory for
Physicists (World Scientific, Singapore, 1989)

\REF{[25]} W. D\"{u}r, G. Vidal, and J. I. Cirac, Phys. Rev. {\bf A62}
 062314 (2000)

\REF{[26]} D. M. Greenberger, M. A. Horne, and A. Zeilinger, in Bell's
Theorem, Quantum Theory, and Conceptions of the Universe,
edited by M. Kafatos (Kluwer, Dordrecht, 1989) p. 69

\newpage
\begin{center}\begin{minipage}{109mm}{\bf Table 1. }
All possible $3$-particle entangled states with $3$ nonzero
terms. There are $24$ one-particle separable states
(Case I) and
$32$ genuine $3$-particle entangled states (Case II).
The symbol $\vert ijk\rangle$ means that the state is
a linear combination of the $i$-th, $j$-th, and $k$-th
basis vectors defined in the text.
\end{minipage}\vskip .2cm{\begin{small}
\begin{tabular}{ccccccccc}
\hline
Case I\\
\hline
$\vert127\rangle$
   &$\vert128\rangle$ &$\vert136\rangle$
   &$\vert138\rangle$ &$\vert146\rangle$&$\vert147\rangle$
   &$\vert167\rangle$&$\vert168\rangle$\\
  $\vert178\rangle$&$\vert235\rangle$&$\vert238\rangle$
  &$\vert245\rangle$
   &$\vert247\rangle$&$\vert257\rangle$
  &$\vert258\rangle$&$\vert278\rangle$ \\
  $\vert345\rangle$&$\vert346\rangle$&$\vert356\rangle$
  &$\vert358\rangle$ &$\vert368\rangle$&$\vert456\rangle$&$\vert457\rangle$
  &$\vert467\rangle$ \\
   \hline
    Case II\\
\hline
   $\vert123\rangle$&$\vert124\rangle$ &$\vert125\rangle$
   &$\vert126\rangle$&$\vert134\rangle$
   &$\vert135\rangle$&$\vert137\rangle$
   &$\vert145\rangle$\\
   $\vert148\rangle$ &$\vert156\rangle$&$\vert157\rangle$ &$\vert158\rangle$
   &$\vert234\rangle$&$\vert236\rangle$&$\vert237\rangle$
   &$\vert246\rangle$\\
   $\vert248\rangle$ & $\vert256\rangle$
   &$\vert267\rangle$ &$\vert268\rangle$
   &$\vert347\rangle$&$\vert348\rangle$&$\vert357\rangle$
   &$\vert367\rangle$\\
   $\vert378\rangle$ & $\vert458\rangle$
   &$\vert468\rangle$ &$\vert478\rangle$
   &$\vert567\rangle$&$\vert568\rangle$&$\vert578\rangle$
   &$\vert678\rangle$\\
   \hline
\end{tabular}\end{small}}
\end{center}
\end{document}